\documentclass [12pt]{iopart}
\usepackage{iopams}
\usepackage{graphicx}
\usepackage{cite}

\def\be{\begin{equation}}
\def\ee{\end{equation}}
\newcommand{\Qp}{{\mathbb Q}_p}
\newcommand{\Zp}{{\mathbb Z}_p}
\newcommand{\Z}{\mathbb Z}
\newcommand{\p}{$p$}
\newcommand{\xip}[1]{\chi_p\left(#1\right)}

\newcommand{\pnorm}[1]{\vert#1\vert_p}
\newcommand{\set}[1]{\left\{#1\right\}}
\renewcommand{\o}{\mathrm{0}}
\newcommand{\I}{\mathrm{I}}
\newcommand{\II}{\mathrm{II}}

\newtheorem{lem}{\emph{Lemma}}
\newtheorem{stat}[lem]{\emph{Proposition}}
\newtheorem{thm}[lem]{\emph{Theorem}}
\newenvironment{proof}{
\par\vspace*{2mm} \emph{\textbf{Proof}}
}{\hspace*{\fill}\nolinebreak[1]\hspace*{\fill}$\square$\par\vspace*{5mm}
}

\begin{document}
\title[Non--Degenerate Ultrametric Diffusion]
{Non Degenerate Ultrametric Diffusion}
\author{S.V. Kozyrev, V.Al. Osipov, V.A. Avetisov}
\address{Semenov Institute of Chemical Physics RAS, Kossygina 4,
119991 Moscow, Russia}
\ead{avetisov@chph.ras.ru}
\begin{abstract}
General non-degenerate \p-adic operators of ultrametric diffusion are introduced.
Bases of eigenvectors for the introduced operators are constructed and the
corresponding eigenvalues are computed. Properties of the corresponding
dynamics (i.e. of the ultrametric diffusion) are investigated.
\end{abstract}
\submitto{JPA}
\maketitle

\section{Introduction}

Ultrametric diffusion models were investigated in relation with models of complex systems,
see for instance \cite{Ogielsky, Blumen}. Mathematical theory of ultrametric diffusion was investigated in
\cite{VVZ, Albeverio, Kochubei}.
In the recent works~\cite{ABK, ABKO, ABO}, \p-adic models of ultrametric
diffusion has been discussed in connection with description of protein
dynamics and characteristic types of relaxation in
complex systems.
Ultrametric diffusion models are naturally related to
the basin--to--basin kinetics approach proposed rather long time ago. Basin--to--basin kinetics approach 
was widely used in computer study of the dynamics constrained by rough multidimensional energy landscapes~\cite{Stillinger, Karplus, Wales, Shaitan, Despa, Shaitan1}.

The basin--to--basin kinetics approach can be outlined as follows.
Let us consider a system, which is described by a particle performing random
walk in a rough energy landscape. The system is supposed to be arriving to the nearest quasi--equilibrium state
(the nearest local minimum on the energy landscape) from any initial
state in the time, which is much smaller compared to the lifetime of this
quasi--equilibrium state.
Therefore, we will reduce our consideration to
the set of local minima of the energy landscape.
Further, we assume that the set of local minima
can be represented as a union of hierarchically nested subsets.
These subsets we will call the basins of minima.
Each of the basins is a union of the non--overlapping basins of
the smaller size (subbasins), each of these smaller basins is a union of the
still smaller ones etc. Moreover, we assume that the larger basins are
separated by the higher activation barriers, and the smaller subbasins
are separated by the lower barriers, i.e. if the basin $A$
is a subbasin of the basin $B$, then the activation barriers between
the maximal subbasins of $A$ is smaller then the activation barriers between
the maximal subbasins of $B$. The basin hierarchy corresponds to the hierarchy of the configuration rearrangements, and the hierarchy of the activation barriers corresponds to the hierarchy of
characteristic times of these rearrangements. 

\begin {figure}
  \centering
  \includegraphics[width=4in, height=4in]{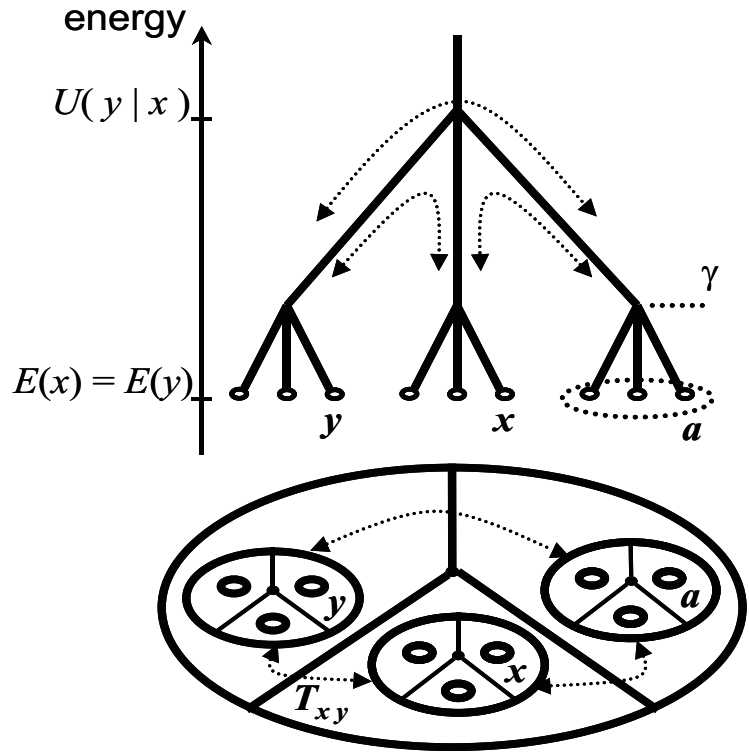}
  \caption {Hierarchy of basins and activation barriers in the basin--to--basin kinetic approach.
For explanation of the notations see the text.
} \label {Fig1}
\end {figure}

As a result, the multidimensional energy surface can be represented by a tree --- a ''skeleton'' of a hierarchical landscape~\cite{Karplus, Wales, Frauenfelder, Despa} (Fig.~\ref{Fig1}). This tree reflects the hierarchy of nestings and it is directed (i.e. it is a tree with a partial order). The vertices of the tree correspond to the basins, 
the partial order describes the ordering of the basins (i.e. the vertex
$A$ is larger than $B$ if the corresponding basin $A$ contains the basin
$B$). The local minima of the landscapes correspond to
minimal vertices with respect to the introduced ordering. In general case for finite or infinite tree
the set of local minima will be described by the absolute of the tree. Absolute of a tree is the set of equivalence classes of decreasing paths in the tree. The equivalence class contains all the decreasing paths, where
each two paths coincide starting from some vertex.

Absolute of any tree has a natural structure of ultrametric space, see~\cite{Serre}. Description of a rough energy landscape in terms of hierarchically nested basins is equivalent to the introduction of ultrametric space of states.
Thus, the basin--to--basin kinetics can be understood as a diffusion (more definitely, as a jump process) in the ultrametric space~\cite{ABK, ABKO, ABO}. The transition probability $T_{xy}$ between the states $y$ and $x$ is determined by the position of the vertex $U(y|x)$ in which the tree branches into the paths going to the points $x$ and $y$, and by the energies $E(x)$ and $E(y)$ of the states $x$ and $y$ (Fig.~\ref{Fig1}). In particular, if all the local minima have the same energy, then $T_{xy}$ is determined only by the vertex $U(y|x)$, i.e. the transition probabilities has the property of permutation symmetry $T_{xy} =T_{yx}$.

The well known example of an ultrametric space is the field of \p-adic numbers
$ \Qp $. In the works~\cite{ABK, ABKO, ABO}, \p-adic description of ultrametric diffusion is introduced in the following way. The system states are parameterized by the \p-adic coordinate $x$,
a basin of states corresponds to the \p-adic disk $B_\gamma (a)$
(Fig.~\ref{Fig1}). The \p-adic disc $B_\gamma (a) $ is a set of all
\p-adic numbers $\set {x: \;\pnorm {x-a} \leq p ^\gamma} $,
for which the \p-adic distance from the disk center $a $ ($a\in \Qp $)
is less or equal than the radius $p ^\gamma $, where $ \gamma $ is an integer
($ \gamma\in\Z $). The parameters $\gamma$ and $a$ distinguish the
\p-adic disks $B_\gamma (a)$.

To describe the system evolution we introduce the probability distribution function  $f(x, t) $ which depends
on the \p-adic coordinate $x$ and the real time $t$: the integral
$$
\int _ {B} f (x, t) d\mu (x)
$$
($d\mu (x) $ is the Haar measure on $ \Qp $) is the probability to find
the system in the set $B$ at time $t$.

The evolution of the function $f (x, t) $ is described by the equation
\be\label{master}
\frac {\partial f (x, t)} {\partial t} =
-\int _ {\Qp} (T _ {yx} f (x, t)-T _ {xy} f (y, t)) d\mu (y)
\ee
This is the master equation for ultrametric diffusion,
and the linear integral operator at the RHS of~(\ref{master})
\be\label{generator}
T f (x) = \int _ {\Qp} (T_{yx} f (x)-T _ {xy} f (y)) d\mu (y)
\ee
is called the operator of ultrametric diffusion.

The non--negative kernel $T_{xy}$ is equal to the rate of transition from the state $y$ to the state $x$
($T_{xy}: \,\Qp\times\Qp\mapsto\mathbb {R}_+$).
We consider the kernels $T_{xy}$, which are locally constant outside any
vicinity of $x=y$. The complex valued function
$g (x) $, defined on $ \Qp $, is called locally constant, if
$$
\forall x\in\Qp\quad\exists \gamma\in\Z\quad \forall z\in\Qp : \pnorm {z} \leq p ^ {\gamma} \Rightarrow g (x+z) =g (x)
$$
The sets, on which the function is constant, are called the sets of the local constancy of the function.

In the simplest case the operators of ultrametric diffusion can be introduced by the kernels $T^\o_{xy} $ satisfying the following conditions:
\begin {description}
\item [(i)]\hspace*{\fill}
\be\label {kernel1}
\eqalign{\mbox {for fixed } y\quad\forall x, z\in\Qp : \pnorm {z} \leq\pnorm {x-y} \Rightarrow T ^\o _ {(x+z) \; y} =T ^\o _ {xy}\\
\mbox {for fixed } x\quad\forall y, z\in\Qp : \pnorm {z} \leq\pnorm {x-y} \Rightarrow T ^\o _ {x \; (y+z)} =T ^\o _ {xy}}
\ee

\item [(ii)]
$$
T ^\o _ {xy} =T ^\o _ {y x}
$$
i.e. the kernel $T ^\o_{xy} $ is symmetric (and
the corresponding operator is Hermitian, since $T ^\o _ {xy} $
is a real--valued function).
\item [(iii)]
$$
\forall a\in\Qp \;\; T ^\o _ {xy}=T ^\o _ {x-a \; y-a}
$$
i.e. the kernel $T ^\o_{xy}$ has the property of translation invariance.
\end {description}

In the following we will denote by $T ^\o$ the operator of ultrametric diffusion, and by $T ^\o_{xy} $ the corresponding kernel.
The general form of the kernels $T ^\o_{xy} $ is given by the series
\be\label{kernel0_1}
T ^\o _ {xy} = \sum _ {\gamma =-\infty} ^ {\infty} T ^ {(\gamma)} \delta _ {p ^\gamma, \pnorm {x-y}}
\ee
where $\delta$ is the Kronecker delta. The kernels $T ^\o$ can be equivalently described by the functions dependent only on \p-adic norm of the difference of $x$ and $y $: $T ^\o _ {x, y} = \rho (\pnorm {x-y})$. The coefficients $T ^ {(\gamma)} $ of the series~(\ref{kernel0_1}) and the function $\rho (\pnorm {x-y}) $ are connected by the
relation $T ^ {(\gamma)} = \rho (p ^\gamma) $.

The properties (i) - (iii) allow us to use the \p-adic Fourier transformation
to compute the eigenvalues of the operator.
If the series $ \sum _ {\gamma=0} ^ \infty p ^\gamma T ^ {(\gamma)}$
converges, the eigenvalues of
the operator $T ^\o $ are determined by the expression
$$
\lambda ^\o_\gamma=p ^ {\gamma} T ^ {(\gamma)} + (1-p ^ {-1}) \sum _ {\gamma ' = \gamma+1} p ^ {\gamma '} T ^ {(\gamma ')}
$$
Every eigenvalue $ \lambda ^\o_\gamma $ is infinitely degenerate.

The operators of the form $T ^\o $ were discussed in the context
of \p-adic mathematical physics. When the kernel has the form
with
$$
T _ {xy} = \frac {p ^\alpha-1} {1-p ^ {-1-\alpha}} \pnorm {x-y} ^ {-(1 +\alpha)} \;, \quad \alpha> 0
$$
the operator $T ^\o $ is the Vladimirov operator of \p-adic
fractional derivation~\cite{VVZ}. Its eigenvalues are given by $ \lambda_\gamma=p ^ {(1-\gamma) \alpha} $, $\gamma\in\Z $. Different examples of the operators $T ^\o $ have been recently investigated in~\cite{ABO}.

In the context of basin--to--basin kinetics, the operator symmetry (hermiticity) property means that all
the local minima of the energy landscape have equal energy. The translation invariance of the kernel means that the transition between $x$ and $y$ depends only on the ultrametric distance between $x$ and $y$.

The translationally invariant operators $T ^\o $ are related to the Parisi matrices (see~\cite{ABK, PaSu}) which were used in the replica approach to spin glasses~\cite{MPV}. However, the energy landscapes of many other disordered systems (for instance, the energy landscapes of clusters, macromolecular structures and biopolymers, see for example~\cite{Wales}) do not have such special properties. Therefore, the ultrametric diffusion operators, different from $T ^\o $, are of great importance.

Generally, the ultrametric diffusion operator $T $ can be defined by~(\ref{generator}) where the kernel satisfies some weaker conditions than conditions (i)--(iii). In the present paper we consider translationally \textit{noninvariant} operators of ultrametric diffusion satisfying the hermiticity property. We will examine two types of such operators, $T^{\I}$ and $T^{\II}$.

A family of operators of \p-adic diffusion $T^{\I}$ has been recently 
investigated in the paper~\cite{Kozyrev}. The local constancy conditions for these operators
are given by~(\ref{kernel1}), as well as for $T ^\o $, but $T ^ {\I}$ differs from $T ^\o$ by violating
the condition (iii).

The kernel $T ^ {\I}_{xy}$ is described by the expression
\be\label {kernel1_1}
T ^\I _ {xy} = \sum _ {\gamma =-\infty} ^ {\infty}
\sum _ {n\in\Qp/\Zp} T ^ {(\gamma n)}
\delta _ {1, \pnorm {p ^\gamma x-p ^\gamma y}} \Omega (p ^\gamma x-n)
\ee
where the factorgroup $ \Qp/\Zp $ is identified with a set of the
fractions $ \sum _ {\gamma=1} ^ k n_\gamma p ^ {-\gamma} $, $n_\gamma=0,
\ldots, p-1 $, $k$ is any natural number, and the coefficients $T ^ {(\gamma n)} \geq 0 $. The function $\Omega(\pnorm {x})$ is an indicator of the \p-adic disk
$$
\Omega (\pnorm {x}) =\cases{ 1,& $\pnorm{x} \le 1$\\0,& $\pnorm{x} > 1$}
$$

It was shown that if the series $ \sum _ {\gamma ' = \gamma} ^ \infty p ^ {\gamma '} T ^ {(\gamma ' n)} $
converges, the eigenvalues of the corresponding operator are given by
\be\label {lambda0}
\lambda ^\I _ {\gamma n} =p ^ {\gamma} T ^ {(\gamma n)} + (1-p ^ {-1}) \sum _ {\gamma ' = \gamma+1} ^ \infty
p ^ {\gamma '} \sum _ {n '\in\Qp/\Zp} T ^ {(\gamma 'n')} \delta _ {n ',\; np ^ {\gamma '}}
\ee
and, in general, are $p-1$ times degenerate.
The eigenvalue $ \lambda ^\I _ {\gamma n} $ corresponds to
the eigenvectors
\be\label{wawelets}
\varphi _ {\gamma n j} (x) =p ^ {-\gamma/2} \xip {p ^ {\gamma-1} jx}
\Omega (\pnorm {p ^ {\gamma} x-n})
\ee
where $j=1, \ldots, p-1 $. Here the function $ \xip {x} = \exp (2\pi i\set {x} _p) $ is an additive
character on the field $ \Qp $, the symbol $ \set {x} _p $ denotes
a fractional part of the \p-adic number. Recall that if the canonical
decomposition of the \p-adic number $x $ is given by
\be\label {canon}
 x = p ^\gamma \sum _ {\mu=0} ^ {\infty} x_\mu p ^\mu,
\quad x_\mu=0, \ldots, p-1, \quad x_0\ne0
\ee
then the fractional part of the number $x $ is determined by
the following expression
$$
\set {x}_p =
\left\{{\begin{array}{ll}0, &\gamma  \ge 0 \\
p^\gamma (x_0 + x_1 +\dots
+ x_{|\gamma| - 1} p^{|\gamma| - 1}), &\gamma <0\end{array}}\right.
$$
As shown in~\cite{Kozyrev1}, the set of vectors~(\ref{wawelets})
forms an orthonormal basis in $L^2 (\Qp) $, which was called
there the basis of \p-adic wavelets.

In the present paper, in section 2, we propose a new general expression for
the kernel $T ^\I $ and show the equivalence of this expression
with~(\ref{kernel1_1}).

In section 3 we introduce the new type of transitionally noninvariant
kernel, $T ^\II $ (more general compared to $T ^\I $), satisfying the weaker
(compared with~({\ref{kernel1}})) condition of local constancy:
\be\label{kernel2}
\eqalign{\mbox{for fixed }y\quad\forall x, z\in\Qp\, :
\pnorm{z}\leq p^{-1}\pnorm{x-y} \Rightarrow T^\II_{(x+z)\;y}=T^\II_{xy}\\
\mbox{for fixed }x\quad\forall y, z\in\Qp\, : \pnorm{z}\leq p^{-1}\pnorm{x-y} \Rightarrow T^\II_{x\;(y+z)}=T^\II_{xy}}
\ee
We introduce the kernel $T ^ {\II} $ both in the functional
form and in the form of a series, and we show the equivalence of
these definitions. In this section, we find the
eigenfunctions of the introduced operator,
which form the basis in $L^2 (\Qp) $, compute the corresponding eigenvalues, and
show that the eigenvalues, in general, are non degenerate.

In section 4, we investigate the properties of ultrametric diffusion,
generated by the operators $T ^\I $ and $T ^\II $.
We consider the relaxation of the initially localized state.
We show that the inhomogeneities of the landscape described by
the translationally noninvariant kernels  $T ^\I $ and $T ^\II $
are not essential for long--time relaxation, i.e. the asymptotics for the relaxation, 
correspondent to kernels $T^{0}_{xy}$, $T^{I}_{xy}$, $T^{II}_{xy}$ will be the same 
for the cases when the corresponding kernels are naturally related in the way described in Section 4..

\section{Translationally noninvariant operators
of ultrametric diffusion $T ^\I $}

\begin {figure}
\centering
  \includegraphics [width=3in, height=2.9in] {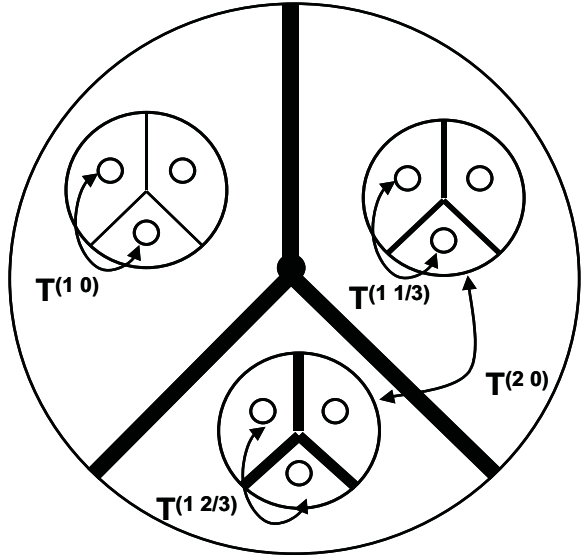}
  \caption {Scheme of the transitions
corresponding to the operator $T ^\I $ with $p=3 $. The lines of different thickness correspond to the different transition probabilities $T ^ {(\gamma n)}$ between the \p-adic disks, which are marked by the circles.
} \label{UnRegBarpicture}
\end {figure}

In the present section we propose a new expression for the kernel $ T ^\I $
and show that this new expression is equivalent to the already known one.
Expression~(\ref{kernel1_1}) implies, that if
$x\in B_\gamma (p ^ {-\gamma} n)$ and $y$ satisfies the condition
$ \pnorm {x-y} =p ^\gamma $, then the kernel $ T ^\I _ {xy} $ is equal to
$T ^ {(\gamma n)} $. In this case, we can illustrate
the basin--to--basin transitions with the help of the scheme on 
Fig.~\ref{UnRegBarpicture}. The given transition scheme allows
to construct the expression of the operator~$ T ^\I $.
To construct a new expression for the kernel  $T ^\I _ {xy} $
we use the following approach. The integration kernel
$T ^\I _ {xy} $ of the operator should depend on the two arguments:
(i) on the size of the minimal disk containing $x $ and $y $,
which is equal to the distance $ \pnorm {x-y} $;
(ii) on the argument distinguishing the disk among the other disks
of the same size. For the last purpose we fix the disk center.
Since any point belonging to an ultrametric disk is its center,
we will take as the center of the minimal disk, containing $x$ and $y$ the following
\begin{equation}\label{fracpart}
\frac{\set {x\pnorm {x-y}} _p} {\pnorm {x-y}}
= \left\{\begin{array}{ll} \sum_{\mu=\log_p \pnorm{x-
y}+1}^{\log_p \pnorm{x} }p^{-\mu}x_\mu\;,&\pnorm{x}>\pnorm{x-y}\\
0,&\pnorm{x}\leq\pnorm{x-y}
\end{array}\right.
\end{equation}
where $x_\mu $ are the coefficients of the  canonical
decomposition~(\ref{canon}) of the \p-adic number $x $.
Therefore, the kernel of the operator $T ^\I $ can be represented by the function
\begin {equation} \label{UnRegKernel}
\rho\left (\pnorm {x-y}, \set {x\pnorm {x-y}} _p\right)
\end {equation}
The function~(\ref{UnRegKernel}) obviously do not have the translational
invariance property. For this function the following proposition is
satisfied.

{\stat The function~(\ref{UnRegKernel}) is symmetric with respect to the
$x\mapsto y$, $y\mapsto x$ permutation.
The function~(\ref{UnRegKernel}) satisfies
the conditions~(\ref{kernel1}) and~(\ref{kernel1_1}).
Moreover, any function satisfying
the condition~(\ref{kernel1_1}) can be represented in the form~(\ref{UnRegKernel}).
Therefore the representations of the kernel $T ^\I $ in the forms~(\ref{UnRegKernel}) and~(\ref{kernel1_1}) are equivalent, and the equivalence is given by the relation
\begin{equation}\label{11.1}
T ^ {(\gamma
n)} = \rho\left (p ^\gamma, n\right) $, $ \gamma\in\Z $, $n\in\Qp/\Zp
\end{equation}
}

\begin {proof} Let us prove the permutation symmetry:
$ {\set {y\pnorm {x-y}} _p = \set {x\pnorm {y-x}} _p} $.
It follows that
$$
\set {x\pnorm {x-y}} _p-\set {y\pnorm {x-y}} _p =
\set {(x-y) \pnorm {x-y}} _p=0
$$

For proving the rest of the proposition, we will show at first that
any function of the form~(\ref{UnRegKernel}) can be represented in
the form~(\ref{kernel1_1}). Actually, for any
$x\in B_\gamma (p ^ {-\gamma} n) $ ($n\in\Qp/\Zp $) and any
$y\in \Qp $ satisfying the condition $ \pnorm {x-y} =p ^\gamma $,
we have $ \set {x\pnorm {x-y}} _p =\set {n} _p=n $. Hence,
$$
\rho\left (\pnorm {x-y}, \set {x\pnorm {x-y}} _p\right) =
\sum _ {\gamma =-\infty} ^ {\infty} \sum _ {n\in\Qp/\Zp}
\rho\left (p ^\gamma, n\right)
\delta _ {1, \pnorm {p ^\gamma x-p ^\gamma y}} \Omega (p ^\gamma x-n)
$$
This means that the function~(\ref{UnRegKernel}) can be represented
in the form of the series~(\ref{kernel1_1}) with the coefficients
$T ^ {(\gamma n)} = \rho\left (p ^\gamma, n\right) $.

On the other hand, since no special restrictions are imposed on
the function~(\ref{UnRegKernel}), then for all $ \gamma\in\Z $ and
$n\in\Qp/\Zp $ we can put $ \rho\left (p ^\gamma, n\right) =
T ^ {(\gamma n)} $. Therefore the representations of the kernel
$T ^\I $ in the forms~(\ref{UnRegKernel}) and~(\ref{kernel1_1})
are equivalent, and the equivalence is given by~(\ref{11.1}).

From the equivalence, it follows that the functions~(\ref{UnRegKernel})
and~(\ref{kernel1_1}) have the same properties. In particular,
the function~(\ref{UnRegKernel}) satisfies the condition~(\ref{kernel1}).
\end{proof}

\section{Translationally noninvariant operators of ultrametric
diffusion $T ^\II $ and the basis of generalized \p-adic wavelets}

Define the family of the operators of ultrametric diffusion
$T ^\II $, with locally constant kernels of the most general form
\be\label {lemma3}
T ^\II _ {xy} = \sum _ {\gamma =
-\infty} ^ {\infty} \sum _ {n\in\Qp/\Zp}
\sum _ {{j, k=0} \atop {k\ne j}} ^ {p-1}
T ^ {(\gamma n jk)} \Omega (p\pnorm {p ^ {\gamma} x-n-j})
\Omega (p\pnorm {p ^ {\gamma} y-n-k})
\ee
where $T ^ {(\gamma n jk)} =T ^ {(\gamma n kj)} \ge 0 $.

{\thm The function of the form~(\ref{lemma3}) is symmetric with respect
to permutation of the arguments, positive, and satisfies the
condition~(\ref{kernel2}).

Moreover, an arbitrary positive symmetric function
satisfying~(\ref{kernel2}) can be represented in the form~(\ref{lemma3}).
}
\begin {proof} Positivity of $T ^\II _ {xy} $ is obvious.
Permutation symmetry is obvious.

Prove that $T ^\II _ {xy} $ given by~(\ref{lemma3})
satisfies~(\ref{kernel2}). This property is easy to check for any
product of two indicator functions in~(\ref{lemma3}).
By linearity, this proves that $T ^\II _ {xy} $ satisfy~(\ref{kernel2}).

Vice versa, it is easy to see that the kernel~(\ref{lemma3})
for $x $, $y $ lying in the disks with the center in
$p ^ {-\gamma} (n+j) $ and $p ^ {-\gamma} (n+k) $ correspondingly and
the radius $p ^ {\gamma-1} $, takes the value $T ^ {(\gamma n jk)} $.

Since all the space $x, y\in \Qp\times \Qp $ is the disjoint union of
such subsets, therefore, taking an arbitrary positive and symmetric
with respect to $j $, $k $ coefficients $T ^ {(\gamma n jk)} $,
we are able to construct an arbitrary symmetric positive
kernel satisfying~(\ref{kernel2}).
This finishes the proof of the theorem.
\end {proof}

The kernel $T ^\II $ can be equivalently described in the functional form.
{\stat The function
\begin{equation} \label{UnRegjKernel}
\rho\left (\pnorm {x-y},
\set {x\pnorm {x-y}} _p, \set {xp ^ {-1} \pnorm {x-y}} _p,
\set {yp ^ {-1} \pnorm {x-y}} _p\right)
\end{equation}
satisfies the condition~(\ref{kernel2}). Moreover, any function satisfying
the condition~(\ref{kernel2}) can be represented in
the form~(\ref{UnRegjKernel}).

Under condition of symmetry of the function~(\ref{UnRegjKernel})
with respect to permutation of $x$ and $y$,
the kernel $T ^\II $ can be equivalently represented
in the forms~(\ref{UnRegjKernel}) and~(\ref{lemma3}),
and the equivalence is given by the relation
\be\label{Fof}
T ^ {(\gamma n jk)} =
\rho\left (p ^ {\gamma}, n, p ^ {-1} (n+j), p ^ {-1} (n+k) \right)
\ee}

\begin {proof}
Let us show first that any function of
the form~(\ref{UnRegjKernel}) can be represented in the form~(\ref{lemma3}).
Actually, any $x $ and $y $ lying in the disks of the radius $p ^ {\gamma-1} $
with the centers in $p ^ {-\gamma} n+p ^ {-\gamma} j $ and
$p ^ {-\gamma} n+p ^ {-\gamma} k $ ($n\in\Qp/\Zp $, $j, k=0, \ldots, p-1 $
and $j\neq k $) can be represented in the form
$$
x=p ^ {-\gamma} (n+j+pz _ {x}), \quad \pnorm {z _ {x}} \leq1
$$
$$
y=p ^ {-\gamma} (n+k+pz _ {y}), \quad \pnorm {z _ {y}} \leq1
$$
Whence it follows that at $j\neq k $, $ \pnorm {x-y} =p ^\gamma $,
$ \set {x\pnorm {x-y}} _p=n $ and
$$
\set {xp ^ {-1} \pnorm {x-y}} _p =
\set {p ^ {-1} (n+j) +z _ {x}} _p=p ^ {-1} (n+j)
$$
$$
\set {yp ^ {-1} \pnorm {x-y}} _p =
\set {p ^ {-1} (n+k) +z _ {y}} _p=p ^ {-1} (n+k)
$$
Hence,
\begin{eqnarray*}
\fl \rho\left (\pnorm {x-y}, \set {x\pnorm {x-y}} _p,
\set {xp ^ {-1} \pnorm {x-y}} _p, \set {yp ^ {-1} \pnorm {x-y}} _p\right) =\\
\sum _ {\gamma =
-\infty} ^ {\infty} \sum _ {n\in\Qp/\Zp}
\rho\left (p ^ {\gamma}, n, p ^ {-1} (n+j), p ^ {-1} (n+k) \right)
\Omega (p\pnorm {p ^ {\gamma} x-n-j}) \Omega (p\pnorm {p ^ {\gamma} y-n-k})
\end{eqnarray*}
Thus, the function~(\ref{UnRegjKernel}) can be represented in the form
of the series~(\ref{lemma3}), whose coefficients are determined by~(\ref{Fof}).

On the other hand, since no restrictions (except for permutation symmetry)
are imposed on the function of the type~(\ref{UnRegjKernel}), then for all
$ \gamma\in\Z $ and $n\in\Qp/\Zp $, $j, k=0, \ldots, p-1 $ ($j\neq k $)
we can put $ \rho\left (p ^ {\gamma}, n, p ^ {-1} (n+j),
p ^ {-1} (n+k) \right) $ to be equal to $T ^ {(\gamma njk)} $.
Therefore the representations of the kernel in the forms
(\ref{UnRegjKernel}) and~(\ref{lemma3}) are equivalent.

Moreover, for the function~(\ref{UnRegjKernel}) and the
series~(\ref{lemma3}) respectively,
the condition of symmetry with respect to permutation of $x$ and $y$
in~(\ref{UnRegjKernel}) and of the symmetry of the coefficients
$T ^ {(\gamma n jk)} $ with respect to permutation of $j$ and $k$,
are equivalent.

From this equivalence, it follows that
the function~(\ref{UnRegjKernel}) satisfies the
condition~(\ref{kernel2}), and any function satisfying the
condition~(\ref{kernel2}) can be represented in the
form~(\ref{UnRegjKernel}).
\end{proof}

\begin {figure}
  \centering
  \includegraphics [width=3in, height=2.9in] {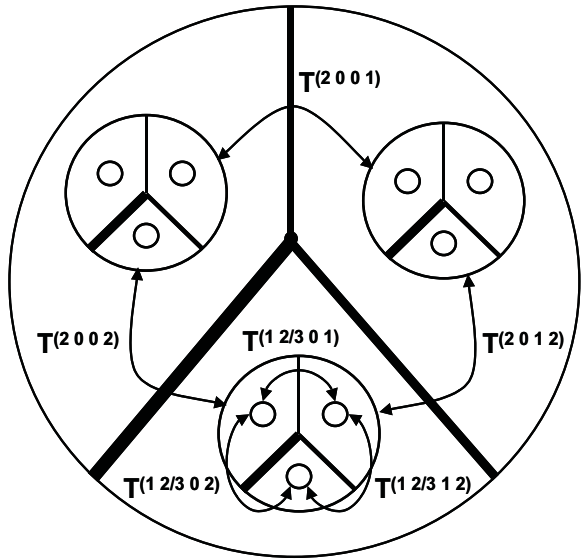}
  \caption {Scheme of the transitions corresponding to the operator $T ^\II $ with $p=3$.
} \label {Diagram_Unregpicture}
\end {figure}

Note that, unlike for the operator $T ^\I $, the kernel of the operator
$T ^\II $ formally depends on a couple of additional functions:
$ \set {xp ^ {-1} \pnorm {x-y}} _p $ and
$ \set {yp ^ {-1} \pnorm {x-y}} _p $. We will explain the meaning of
these functions. Consider the transition between the states $x $ and $y $
(see fig.~\ref{Diagram_Unregpicture}). The probability for $x \mapsto y $
transition depends on a relative position of the basins between which
the transition is carried out. Let $x $ and $y $ belong to the basin described
by the \p-adic disk $B_\gamma (a) $ of the radius $p ^\gamma =\pnorm {x-y} $
with the center in $a =\set {x p ^\gamma} _pp ^ {-\gamma} $.
Then the radii of the disks $B _ {\gamma-1} (b) $ and $B _ {\gamma-1} (c) $,
between which the transition is carried out, are equal to
$p ^ {-1} \pnorm {x-y} $, and the disk centers are determined by
the functions $b =\set {x p ^ {\gamma-1}} _p p ^ {-\gamma+1} $,
$c =\set {y p ^ {\gamma-1}} _p p ^ {-\gamma+1} $.

{\stat \label {peq2} If $T ^ {(\gamma n jk)} $ in~(\ref{lemma3})
doesn't depend on $j $ and $k $, then the operator $T ^\II $
reduces to $T ^\I $.}

\begin {proof}
Consider the kernel $T ^\II _ {xy} $ under the condition that
$T ^ {(\gamma n jk)} $ is independent on $j $ and $k $:
\be\label {vvv}
T ^\II _ {xy} = \sum _ {\gamma, n} T ^ {(\gamma n)}
\sum _ {j=0} ^ {p-1} \Omega (p\pnorm {p ^ {\gamma} x-n-
j}) \sum _ {{k=0} \atop {k\ne j}} ^ {p-1}
\Omega (p\pnorm {p ^ {\gamma} y-n-k})
\ee
Using the formula
\begin{eqnarray}\label{formula}
\fl \Omega (\pnorm {p ^ {\gamma} x-n}) =
\Omega (p ^ {-\gamma} \pnorm {x-p ^ {-\gamma} n}) \nonumber\\=
\sum _ {j=0} ^ {p- 1} \Omega (p ^ {-\gamma+1}
\pnorm {x-p ^ {-\gamma} n-p ^ {-\gamma} j}) =
\sum _ {j=0} ^ {p-1} \Omega (p\pnorm {p ^ {\gamma} x-n-j})
\end{eqnarray}
as well as the property of indicators of the disks
$$
\Omega (\pnorm {x-a}) \Omega (\pnorm {y-a}) =
\Omega (\pnorm {x-a}) \Omega (\pnorm {x-y})
$$
for the sum on $j $ and $k $ in~(\ref{vvv}) we have
\begin{eqnarray*}
\fl\sum _ {j=0} ^ {p-1} \Omega (p\pnorm {p ^ {\gamma} x-n-j}) \sum _ {{k=0} \atop {k\ne j}} ^ {p-1} \Omega (p\pnorm {p ^ {\gamma} y-n-k})
\\ \lo =\sum_{j=0}^{p-1} \Omega (p\pnorm {p ^ {\gamma} x-n-j}) (\Omega (\pnorm {p ^ {\gamma} y-n})-\Omega (p\pnorm {p ^ {\gamma} y-n-j}))
\\ \lo =\Omega (\pnorm {p ^ {\gamma} x-n}) (\Omega (\pnorm {p ^ {\gamma} y-n})-\Omega (p\pnorm {p ^ {\gamma} y-p ^ {\gamma} x}))=\Omega (\pnorm {p ^ {\gamma} x-n})  \\
\times(\Omega (\pnorm {p ^ {\gamma} y-p ^ {\gamma} x})-\Omega (p\pnorm {p ^ {\gamma} y-p ^ {\gamma} x})) = \Omega (\pnorm {p ^ {\gamma} x-n}) \delta _ {1, \pnorm {p ^\gamma x-p ^\gamma y}}
\end{eqnarray*}
\end {proof}
Note that the operators $T ^\I $ and $T ^\II$ are identically equal for $p=2$.
Actually, in this case the expression~(\ref{lemma3}) contains only
$T^{(\gamma n 01)} $ and $T^{(\gamma n 10)}$.
Since $T ^ {(\gamma n 01)} =T ^ {(\gamma n 10)}$,
from the proposition~\ref{peq2} it follows that
$T ^\I _ {xy} =T ^\II _ {xy} $.

Now we will construct the basis of eigenfunctions of the operator $T ^\II $.

Consider the $p\times p $ matrix
$ \left ({\bf W} ^ {(\gamma n)} \right) _ {ls} $ with matrix elements equal
to $-T ^ {(\gamma n jk)} $ for $j\ne k $ and equal to
$ \sum _ {k=0\atop {k\ne l}} ^ {p-1} T ^ {(\gamma n kl)} $
for the diagonal elements:
\be\label {bfT}
{\bf W} ^ {(\gamma n)} _ {ls} =
\delta _ {sl} \sum _ {k=0\atop k\ne l} ^ {p-1} T ^ {(\gamma n lk)} -
(1-\delta _ {sl}) T ^ {(\gamma n ls)}
\ee
It is easy to see that $ {\bf W} ^ {(\gamma n)} $ is a real symmetric
$p\times p $ matrix. Moreover, the matrix is positive:
{\lem The matrix $ {\bf W} ^ {(\gamma n)} $ defined by~(\ref{bfT})
is positive.}
\begin {proof}
Compute the Hermitian form (we omit the $ {(\gamma n)} $ index)
$$
\langle z, {\bf W} z\rangle = \sum _ {s=0} ^ {p-1} |z_s | ^ 2\sum _ {l=0 \atop l\ne s} ^ {p-1} W _ {sl} - \sum _ {s=0} ^ {p-
1} \sum _ {l=0 \atop l\ne s} ^ {p-1} z ^ * _ s z_l W _ {sl} = \sum _ {s=0} ^ {p-1} \sum _ {l=0 \atop l\ne s} ^ {p-1} W _ {sl}
(|z_s | ^ 2-z ^ * _ s z_l)
$$
Combining the terms containing $W _ {sl} $ and $W _ {ls} $ and using that
$T ^ {(ls)} $ is a real symmetric matrix with nonnegative entries,
we obtain for the combination of the terms
$$
T _ {sl} (|z_s | ^ 2 + | z_l | ^ 2-z ^ * _ s z_l-z ^ * _ l z_s) \ge 0
$$
that proves the positivity and finishes the proof of the lemma.
\end {proof}

The matrix ${\bf W}^{(\gamma n)}$ has $p$ nonnegative eigenvalues.
Let us assume that the $j$-th ($j=1, \dots, p-1 $) eigenvector has the coordinates $h _ {\gamma nj} ^ {k} $, $k=0, \dots, p-1 $ and corresponds to the eigenvalue $\lambda_{j}^{(\gamma n)}$:
$$
\sum _ {k=0} ^ {p-1} W ^ {(\gamma n)} _ {lk} h _ {\gamma n j} ^ {k} =\lambda _ {j} ^ {(\gamma n)} h _ {\gamma n j} ^ {l}
$$
It is easy to see that the matrix ${\bf W}^{(\gamma n)}$ has the zero eigenvalue, which corresponds to the eigenvector with equal matrix elements: $h _ {\gamma n 0} ^ {k} =p ^ {-1/2} $ for all $k=0, \dots, p-1 $.
Since eigenvectors are orthonormal we have
$$
\sum _ {k=0} ^ {p-1} h _ {\gamma n j} ^ {*k} h _ {\gamma n j '} ^ {k} = \delta _ {jj '} \;, \quad j, j ' =0, \dots, p-1
$$

Let us represent all eigenvectors of the matrix $ \mathbf {W} ^ {(\gamma n)} $ in the form of the discrete Fourier  expansion:
\be\label{decompr}
h _ {\gamma n j} ^k =\sum _ {s=0} ^ {p-1} \xip {p ^ {-1} k (n+s)} g _ {\gamma n j} ^s \;,\quad j=0, \ldots, p-1
\ee
From the properties of the discrete Fourier transformation and the properties of eigenvectors of the matrix $ {\bf W} ^ {(\gamma n)} $, it follows that
\be\label {v1}
g _ {\gamma n j} ^s = \sum _ {k=0} ^ {p-1} h _ {\gamma n j} ^k \xip {-p ^ {-1} k (n+s)}
\ee
\be\label {v2}
g _ {\gamma n j} ^0=g _ {\gamma n 0} ^j=0 \;, \quad j=0, \ldots, p-1
\ee
\be\label {v3}
\sum _ {s=1} ^ {p-1} g _ {\gamma n j} ^sg _ {\gamma n j '} ^ {s *} =\sum _ {k=0} ^ {p-1} h _ {\gamma n j} ^k h _ {\gamma n j '} ^ {k *} =\delta _ {jj '}
\ee

Consider the function $ \psi _ {\gamma nj} (x) $ of the form
\begin{eqnarray}
\psi _ {\gamma n j} (x) =p ^ {{1-\gamma} \over 2}
\sum _ {k=0} ^ {p-1} h _ {\gamma n j} ^ {k}
\Omega (p\pnorm {p ^ {\gamma} x-n-k}) \label{wavelet}\\
\gamma\in \Z \;,\quad n\in\Qp/\Zp \;,\quad j=1, \ldots, p-1\nonumber
\end{eqnarray}
Note that $ \psi _ {\gamma n j} (x) $ is a locally constant function, constant on disks of the radius $p ^ {\gamma-1} $ and $ \psi _ {\gamma nj} (x) \in L^2 (\Qp) $. The functions $ \psi _ {\gamma n j} (x) $ we will call the generalized $p$--adic wavelets.

Using the expansion~(\ref{decompr}), the relations~(\ref{v1}),~(\ref{v2}),~(\ref{v3}) and the formula~(\ref{formula}), for $ \psi _ {\gamma n j} (x) $, we have
\begin{eqnarray*}
\fl \psi _ {\gamma n j} (x) =p ^ {-{\gamma/2}} \sum _ {k=0} ^ {p-1} \sum _ {s=1} ^ {p-1} \xip {p ^ {-1} s (n+k)} g _ {\gamma nj} ^s \Omega (p\pnorm {p ^ {\gamma} x-n-k}) \\
\lo =p ^ {-{\gamma/2}} \sum _ {s=1} ^ {p-1} \xip {p ^ {\gamma-1} sx} g _ {\gamma n j} ^s \sum _ {k=0} ^ {p-1} \Omega (p\pnorm {p ^ {\gamma} x-n-k}) \\
= \sum _ {s=1} ^ {p-1} g _ {\gamma n j} ^s p ^ {-{\gamma/2}} \xip {p ^ {\gamma-1} sx}
\Omega (\pnorm {p ^ {\gamma} x-n}) = \sum _ {s=1} ^ {p-1} g _ {\gamma n j} ^s \varphi _ {\gamma n s} (x)
\end{eqnarray*}
Thus, each of the functions~(\ref{decompr}) can be represented in the form of linear combination of \p-adic wavelets~(\ref{wawelets}):
\be\label{rotwavelet}
\psi _ {\gamma n j} (x) = \sum _ {s=1} ^ {p-1} g _ {\gamma n j} ^s \varphi _ {\gamma n s} (x)
\ee

For further computations we use the following identity
\begin{eqnarray}\label {productinds}
\fl\Omega (p\pnorm {p ^ {\gamma} x-n-k}) \Omega (p\pnorm {p ^ {\gamma '} x-n '-l})\nonumber\\
\lo = \delta _ {p ^ {\gamma '-\gamma} (n+k), n ' +l} \theta (\gamma '-\gamma) \Omega (p\pnorm {p ^ {\gamma} x-n-k}) \nonumber\\
 + \delta _ {n+k, p ^ {\gamma-\gamma '} (n ' +l)}(1-\theta (\gamma '-\gamma)) \Omega (p\pnorm {p ^ {\gamma '} x-n'-l})
\end{eqnarray}
where
$$
\theta (\gamma) = \cases{1, &$\gamma> 0$\\
0, &$\gamma\leq 0$}
$$
and $ \delta _ {p ^ {\gamma '-\gamma} (n+k), n ' +l} $
is the Kronecker symbol on the group $ \Qp/p\Zp $:
$$
\Omega (\pnorm {p ^ {\gamma '-\gamma} (n+k) - (n ' +l)}) =
\delta _ {p ^ {\gamma '-\gamma} (n+k), n ' +l} \qquad\mbox {for} \quad
\gamma '\ge \gamma
$$

{\thm The set of functions $ \{\psi _ {\gamma n j} (x) \} $,
$ \gamma\in \mathbb {Z} $, $n\in \Qp/\Zp $, $j=1, \dots, p-1 $
is an orthonormal basis in $L^2 (\Qp) $.}

\begin {proof}
Consider the scalar product
\begin{eqnarray*}
p^{\gamma-1\over 2} p^{\gamma'-1\over 2}
\langle \psi_{\gamma n j},\psi_{\gamma' n' j'}\rangle
\\
=\int_{Q_p}
\sum_{s=0}^{p-1} h_{\gamma n j}^{*s}
\Omega(p|p^{\gamma}x-n-s|_p)
\sum_{l=0}^{p-1} h_{\gamma' n' j'}^{l}
\Omega(p|p^{\gamma'}x-n'-l|_p)
d\mu(x)
\end{eqnarray*}
Using~(\ref{productinds}) we compute the following
\begin{eqnarray*}
\fl p^{\gamma-1\over 2} p^{\gamma'-1\over 2}
\langle \psi_{\gamma n j},\psi_{\gamma' n' j'}\rangle=
\sum_{s=0}^{p-1} h_{\gamma n j}^{*s}
\sum_{l=0}^{p-1} h_{\gamma' n' j'}^{l}
\int_{Q_p}\biggl(\delta_{p^{\gamma'-\gamma}(n+s),n'+l}
\theta(\gamma'-\gamma)\Omega(p|p^{\gamma}x-n-s|_p)
\\\lo
+\delta_{n+s,p^{\gamma-\gamma'}(n'+l)}
(1-\theta(\gamma'-\gamma))\Omega(p|p^{\gamma'}x-n'-l|_p)
\biggr) d\mu(x)
\\\lo =\sum_{s=0}^{p-1} h_{\gamma n j}^{*s}
\sum_{l=0}^{p-1} h_{\gamma' n' j'}^{l}
\biggl(\delta_{p^{\gamma'-\gamma}(n+s),n'+l}
\theta(\gamma'-\gamma)p^{\gamma-1} +\delta_{n+s,p^{\gamma-\gamma'}(n'+l)}
(1-\theta(\gamma'-\gamma))p^{\gamma'-1}
\biggr)
\end{eqnarray*}
In this expression, if the first term is non zero and $\gamma'>\gamma$,
then $\delta_{p^{\gamma'-\gamma}(n+s),n'+l}$ does not depend on $s$
and we obtain the summation on $s$ of the form
$\sum_{s=0}^{p-1} h_{\gamma n j}^{*s}$ which is equal to zero for $j>0$.

In the same way, we prove that the second term does not vanish only for
$\gamma'=\gamma$. This proves that
$$
p^{\gamma-1\over 2} p^{\gamma'-1\over 2}
\langle \psi_{\gamma n j},\psi_{\gamma' n' j'}\rangle=
\delta_{\gamma\gamma'}\delta_{nn'}p^{\gamma-1}
\sum_{s=0}^{p-1} h_{\gamma n j}^{*s}h_{\gamma n j'}^{s}
=\delta_{\gamma\gamma'}\delta_{nn'}\delta_{jj'}p^{\gamma-1}
\|h_{(\gamma n) j}\|^2
$$
which implies that $\{\psi_{\gamma n j}\}$ is an orthonormal system
of functions.

To prove that the set of vectors $ \set {\psi _ {\gamma n j}} $ is
an orthonormal basis (is total in $L^2 (\Qp) $) we use the Parsevale identity.
Since the set of indicators (characteristic functions) of \p-adic disks is
total in $L^2 (\Qp) $ it is enough to check the Parsevale identity for
the indicator $ \Omega (p\pnorm {p ^ {\gamma} x-n-s}) $.
We have for the scalar product of the indicator and the wavelet
$$
\langle \Omega (p\pnorm {p ^ {\gamma} x-n-s}), \psi _ {\gamma 'n' j '} \rangle =p ^ {{1-\gamma '} \over 2} \sum _ {l=0} ^ {p-1} h _ {\gamma ' n 'j'} ^ {l} p ^ {\gamma-1} \biggl (\delta _ {p ^ {\gamma '-\gamma} n, n ' +l}
\theta (\gamma '-\gamma) + \delta _ {\gamma\gamma '} \delta _ {nn '} \delta _ {sl} \biggr)
$$
Summing up the wavelets we get
\begin{eqnarray}\label {parsevale}
\fl\sum _ {\gamma 'n' j '}
| \langle \Omega (p\pnorm {p ^ {\gamma} x-n-s}), \psi _ {\gamma 'n' j '} (x) \rangle | ^ 2\nonumber \\
\lo =p ^ {\gamma-1} \left [\sum _ {j '} |h _ {\gamma n j '} ^ {s} | ^2 + p ^ {\gamma-1}
\sum _ {\gamma '> \gamma; n ' j '} p ^ {1-\gamma '} \sum _ {l=0} ^ {p-1}
|h _ {\gamma 'n' j '} ^ {l} | ^2
\delta _ {p ^ {\gamma '-\gamma} n, n ' +l} \right]
\end{eqnarray}


Using the normalization condition we get
\be\label{26a}
\sum _ {j=1} ^ {p-1} h _ {\gamma n j} ^ {* s} h _ {\gamma n j} ^ {s} =1-p ^ {-1}
\ee
which implies for~(\ref{parsevale})
$$
(1-p ^ {-1}) p ^ {\gamma-1} \left [1 + p ^ {\gamma-1}
\sum _ {\gamma '> \gamma} p ^ {1-\gamma '}
\right] = (1-p ^ {-1}) p ^ {\gamma-1} (1-p ^ {-1}) ^ {-1} =p ^ {\gamma-1}
$$
that proves the Parsevale identity.
\end {proof}

In the next theorem we prove that the constructed in the theorem above
basis is an eigenbasis of the ultrametric
diffusion operator $T^\II$ and compute the corresponding eigenvalues.

{\thm
Let the kernel~(\ref{lemma3}) satisfies the condition of convergence of the series
\be\label {convergence}
\sum _ {\gamma> 0} \sum _ {k=1} ^ {p-1} p ^ {\gamma} T ^ {(\gamma 0 k 0)}
\ee

Then the operator~(\ref{generator}) is a well defined operator in
$L^2 (\Qp) $ with the dense domain and the generalized
\p-adic wavelets $ \psi _ {\gamma n j} $ are eigenvectors for the
operator $T ^\II$:
$$
T ^\II\psi _ {\gamma nj} = \lambda ^\II _ {\gamma nj} \psi _ {\gamma nj}
$$
with the eigenvalues
\be\label {lambdagn}
\lambda ^\II _ {\gamma nj} = p ^ {\gamma-1} \lambda _ {j} ^ {(\gamma n)} + \sum _ {\gamma 'n' l k\atop l\ne k} p ^ {\gamma '-
1} T ^ {(\gamma 'n' lk)} \delta _ {p ^ {\gamma '-\gamma} n, n ' +l} \theta (\gamma '-\gamma)
\ee}

\begin {proof}
Consider the action of the operator on the wavelet $ \psi _ {\gamma j n} $.
\begin{eqnarray*}\fl
p ^ {\gamma-1\over 2}
T ^\II\psi _ {\gamma nj} (x) =
p ^ {\gamma-1\over 2} \int T ^\II _ {xy} \left (\psi _ {\gamma nj} (x)-\psi _ {\gamma nj} (y)
\right) d\mu (y) \\
\lo = p ^ {\gamma '-1} \sum _ {\gamma ' n'lk\atop l\ne k} T ^ {(\gamma 'n' lk)} \Omega (p\pnorm {p ^ {\gamma '} x-n '-l}) \sum _ {s=0} ^ {p-1} h _ {\gamma n j} ^ {s}
\Omega (p\pnorm {p ^ {\gamma} x-n-s})\\\lo
-\sum _ {\gamma ' n'lk\atop l\ne k} T ^ {(\gamma 'n' lk)} \Omega (p\pnorm {p ^ {\gamma '} x-n '-l}) \sum _ {s=0} ^ {p-1} h _ {\gamma n j} ^ {s} \\\times\int \Omega (p\pnorm {p ^ {\gamma '} y-n '-k}) \Omega (p\pnorm {p ^ {\gamma} y-n-s}) d\mu (y)
\end{eqnarray*}

Using~(\ref{productinds}) for $p ^ {\gamma-1\over 2} T ^\II\psi _ {\gamma nj} (x) $, we compute the following
\begin{eqnarray*}\fl
\sum _ {\gamma ' n'lk\atop l\ne k} T ^ {(\gamma 'n' lk)}
\sum _ {s=0} ^ {p-1} h _ {\gamma n j} ^ {s}
\biggl [
\delta _ {p ^ {\gamma '-\gamma} (n+s), n ' +l}
\theta (\gamma '-\gamma) p ^ {\gamma '-1} \Omega (p\pnorm {p ^ {\gamma} x-n-s}) \\ \lo
+ \left [\delta _ {n+s, p ^ {\gamma-\gamma '} (n ' +l)}
-\delta _ {n+s, p ^ {\gamma-\gamma '} (n ' +k)} \right]
(1-\theta (\gamma '-\gamma)) p ^ {\gamma '-1} \Omega (p\pnorm {p ^ {\gamma '} x-n '-l})\\
-\delta _ {p ^ {\gamma '-\gamma} (n+s), n ' +k}
\theta (\gamma '-\gamma) p ^ {\gamma-1} \Omega (p\pnorm {p ^ {\gamma '} x-n '-l})
\biggr]
\end{eqnarray*}

We prove that the term proportional to $1-\theta (\gamma '-\gamma) $ is equal to the following
$$
\left [\delta _ {n+s, p ^ {\gamma-\gamma '} (n ' +l)}
-\delta _ {n+s, p ^ {\gamma-\gamma '} (n ' +k)} \right]
(1-\theta (\gamma '-\gamma)) = \delta _ {\gamma\gamma '} \delta _ {nn '}
(\delta _ {sl}-\delta _ {sk})
$$
and
$$
\sum _ {s=0} ^ {p-1} h _ {\gamma n j} ^ {s}
\delta _ {p ^ {\gamma '-\gamma} (n+s), n ' +k} \theta (\gamma '-\gamma) =
\delta _ {p ^ {\gamma '-\gamma} n, n ' +k} \theta (\gamma '-\gamma)
\sum _ {s=0} ^ {p-1} h _ {\gamma n j} ^ {s} =0
$$
since for $j=1, \dots, p-1 $, we have $ \sum _ {s=0} ^ {p-1} h _ {\gamma n j} ^ {s} =0 $.

This implies for $p ^ {\gamma-1\over 2} T ^\II\psi _ {\gamma nj} (x) $ the following
\begin{eqnarray*}\fl
p ^ {\gamma-1\over 2} T ^\II\psi _ {\gamma nj} (x) = \left [\sum _ {\gamma ' n'lk\atop l\ne k} T ^ {(\gamma 'n' lk)}
\delta _ {p ^ {\gamma '-\gamma} n, n ' +l} \theta (\gamma '-\gamma) p ^ {\gamma '-1} \right]
\sum _ {s=0} ^ {p-1} h _ {\gamma n j} ^ {s} \Omega (p\pnorm {p ^ {\gamma} x-n-s}) \\+ p ^ {\gamma-1} \sum _ {s=0} ^ {p-1} h _ {\gamma n j} ^ {s} \sum _ {lk\atop l\ne k} T ^ {(\gamma n lk)}(\delta _ {sl}-\delta _ {sk}) \Omega (p\pnorm {p ^ {\gamma} x-n-l})
\end{eqnarray*}
Consider
\begin{eqnarray*}\fl
\sum _ {s=0} ^ {p-1} h _ {\gamma n j} ^ {s}
\sum _ {lk\atop l\ne k} T ^ {(\gamma n lk)}
(\delta _ {sl}-\delta _ {sk})
\Omega (p\pnorm {p ^ {\gamma} x-n-l}) \\\lo
= \sum _ {l=0} ^ {p-1} \Omega (p\pnorm {p ^ {\gamma} x-n-l}) \sum _ {s=0} ^ {p-1}
\left (\delta _ {sl} \sum _ {lk\atop l\ne k} T ^ {(\gamma n lk)}-
(1-\delta _ {sl}) T ^ {(\gamma n ls)} \right)
h _ {\gamma n j} ^ {s} \\
= \sum _ {l=0} ^ {p-1} \Omega (p\pnorm {p ^ {\gamma} x-n-l})
\lambda _ {j} ^ {(\gamma n)}
h _ {\gamma n j} ^ {l} =
p ^ {\gamma-1\over 2} \lambda _ {j} ^ {(\gamma n)} \psi _ {\gamma n j} (x)
\end{eqnarray*}

Finally, we obtain
$$
T ^\II\psi _ {\gamma nj} (x) =\left [p ^ {\gamma-1} \lambda _ {j} ^ {(\gamma n)} + \sum _ {\gamma ' n'lk\atop l\ne k} p ^ {\gamma '-1} T ^ {(\gamma 'n' lk)}\delta _ {p ^ {\gamma '-\gamma} n, n ' +l}\theta (\gamma '-\gamma) \right] \psi _ {\gamma nj} (x)
$$

Using the condition of convergence of the series~(\ref{convergence}) we obtain the proof of the theorem.
\end {proof}

\section{Relaxation problem}
Let us consider the relaxation problem formulated analogously to that in~\cite{ABK, ABKO, ABO} (see also references therein). In these works, the evolution of probability distribution was described by the equation of the
form~(\ref{master}) with the ultrametric diffusion operator of the type $T ^\o $. The initial distribution was taken homogeneous on $\Zp$ (i.e. $f (x, 0) = \Omega (\pnorm {x}) $). 

Consider the relaxation function of the system, $R(t)$,  which describes  the evolution of population of the system in the set where the initial distribution was concentrated. In the case when the initial distribution is the characteristic function of the unit ball with the center in zero, this reduces to 
$$
R (t)=\left<\Omega (\pnorm {x}),e^{-Tt}\Omega (\pnorm {x})\right> = \int _ {\Zp} f (x, t) d\mu (x)
$$

It is known (see, in particular,~\cite{Metzler, Ogielsky, ABK, ABKO, ABO}), for the case when the ultrametric
diffusion is generated by $T ^\o $, the relaxation function $R(t)$
takes the form
\be\label {relax0_1}
R ^\o (t)=\left<\Omega (\pnorm {x}),e^{-T^\o t}\Omega (\pnorm {x})\right> = (p-1) \sum _ {\gamma=1} ^ \infty
p ^ {-\gamma} \exp (-\lambda ^\mathrm {0} _ \gamma t)
\ee

Let us investigate the relaxation behavior for the cases,
when the ultrametric diffusion is generated by the operators $T ^\I $
and $T^\II $. The initial distribution we will take to be equal to the characteristic function of the disk:
\be\label{InitCond}
f (x, 0) = \Omega (\pnorm {x-a}), \quad\pnorm {a}=p^N, \quad N\ge 1
\ee
and consider the relaxation function 
$$
R ^\I (t)=\left<\Omega (\pnorm {x-a}),e^{-T^\I t}\Omega (\pnorm {x-a})\right>
$$
and $R ^\II$, defined analogously.
 
\subsection*{The operator $T^\I $}

Find the coefficients of the decomposition of the initial condition~(\ref{InitCond}) in the basis of $p$--adic wavelets~(\ref{wawelets}).
$$
C _ {\gamma nj} =\left<\Omega (\pnorm {x-a}),\varphi _ {\gamma nj} (x)\right> = \varphi _ {\gamma nj} (a) \theta (\gamma)
$$ 
The solution of the Cauchy problem for the ultrametric diffusion equation~(\ref{master}) with the operator $T ^\I $ and the initial
condition~(\ref{InitCond}) takes the form
$$
f^\I (x, t) = \sum _ {\gamma=1} ^ {\infty} \sum _ {n\in\Qp/\Zp} \sum _ {j=1} ^ {p-1} e ^ {-\lambda ^\I _ {\gamma n} t}\varphi _ {\gamma nj} (a) \varphi ^ * _ {\gamma nj} (x)
$$
Then the relaxation function $R ^\I (t) $ is given by the expression
\begin{eqnarray*}
R ^\I (t) = \sum _ {\gamma=1} ^ {\infty} \sum _ {n\in\Qp/\Zp}
\sum _ {j=1} ^ {p-1} e ^ {-\lambda ^\I _ {\gamma n} t} | \varphi _ {\gamma nj} (a) | ^2 \\=
(p-1) \sum_{\gamma=1} ^ {\infty} \sum _ {n\in\Qp/\Zp} p ^ {-\gamma} e ^ {-\lambda ^\I _ {\gamma n} t} \Omega (\pnorm {p^{\gamma} a-n})
\end{eqnarray*}
Taking into account that $\pnorm{a}=p^N$ and $N\ge 1$, let us divide the series into the two parts: for $\gamma\le N$ and for $\gamma>N$. For the second part the values of all the indicators are equal to one only if $n=0$, otherwise they are equal to zero. Hence, for the relaxation function $R ^\I(t)$ we get   
\be\label{RelaxSerieI}
R ^\I (t) = (p-1) \sum _ {\gamma=1} ^ {N} p ^ {-\gamma} e ^ {-t\lambda ^\I_{\gamma,\;
p^{\gamma}\:a} } + (p-1) \sum _ {\gamma=N+1} ^ \infty p ^ {-\gamma} e ^ {-t\lambda ^\I _ {\gamma 0} t}
\ee

\subsection*{The operator $T ^\II $}

The solution of the Cauchy problem for the ultrametric diffusion
equation~(\ref{master}) with the operator $T ^\II $ and the initial
condition~(\ref{InitCond}) takes the form
$$
f^\II (x, t) = \sum _ {\gamma=1} ^ {\infty} \sum _ {n\in\Qp/\Zp} \sum _ {j=1} ^ {p-1} e ^ {-\lambda ^\II _ {\gamma n j} t} \psi _ {\gamma nj} ^ * (x) \psi _ {\gamma nj} (a)
$$
Then the relaxation function $R ^\II (t) $ is given by the expression
\begin{eqnarray*}
\fl R ^\II (t) =  \sum _ {\gamma=1} ^ {\infty} \sum _ {n\in\Qp/\Zp} \sum _ {j=1} ^ {p-1} e ^ {-\lambda ^\II _ {\gamma n j}} | \psi _ {\gamma nj} (a) | ^2\\ =\sum _ {\gamma=1} ^ {\infty} \sum _ {n\in\Qp/\Zp} \sum _ {j=1} ^ {p-1} e ^ {-\lambda ^\II _ {\gamma n j}} p ^ {1-\gamma} \sum _ {k=0} ^ {p-1} |h _ {\gamma nj} ^ {k} | ^2 \Omega (p\pnorm {p ^ {\gamma} a-n-k})
\end{eqnarray*}
Taking into account that $\pnorm{a}=p^N$ and $N\ge 1$, for the relaxation function $R ^\II(t)$ we get 
\begin{equation*}
\fl R ^\II (t) = \sum _ {\gamma=1} ^ {N} \sum _ {j=1} ^ {p-1} \sum _ {k=0} ^ {p-1} e ^ {-\lambda ^\II _ {\gamma n j} t} p ^ {1-\gamma} |h _ {\gamma nj} ^ {k} | ^2 \delta_{p ^ {\gamma} a-k,\;n} + \sum _ {\gamma=N+1} ^ {\infty} p ^ {-\gamma} \sum _ {j=1} ^ {p-1} p|h _ {\gamma 0j} ^ {0} | ^2 e ^ {-\lambda ^\II _ {\gamma 0j} t}
\end{equation*}

\nosections
Comparing the formulas~(\ref{relax0_1}) and~(\ref{RelaxSerieI}), we see that the long--time relaxation behavior for ultrametric diffusions generated by the operators $T^{0}$ and $T^{I}$ coincide (i.e. the functions $R^{\o}(t)$ and $R^{\I}(t)$ have the same asymptotic). This shows that the particular properties of the energy landscape, such as local inhomogeneities, are not important for long time behavior of the corresponding diffusion. Note that the given result generalizes the special case considered in the work~\cite{Yoshino} by Yoshino.

For the cases of the operators of the type $T ^\II $, the relaxation function behavior becomes more complicated. Note that if the landscape deviations from the regularity are small:
$$
| \lambda ^\II _ {\gamma 0j}-\langle\lambda ^\II _ {\gamma 0j} \rangle_j |\ll\langle\lambda ^\II _ {\gamma 0j} \rangle_j
$$
where $ \langle\lambda ^\II _ {\gamma 0j} \rangle_j = (p-1) ^ {-1} \sum _ {j=1} ^ {p-1} \lambda ^\II _ {\gamma 0j} $, then by~(\ref{26a}) the long-time relaxation $R ^{\I} (t) $ is a good approximation of the relaxation $R ^\II (t)$.

\ack
The authors are grateful to Dr A Kh Bikulov for active discussions. We would like to thank NPO ITIN for technical support. This work has been partially supported by the Dynasty Foundation. S.K. was partially supported by the CRDF (grant UM1--2421--KV--02), the RFBR grant 02--01--01084 and the grant for support of scientific school N.Sh.--1542.2003.1.

\end{document}